# Spot Scan Probe of Lateral Field Effects in a Thick Fully-Depleted CCD

**Paul O'Connor**[a*]

[a] *Brookhaven National Laboratory*
*Upton, NY USA 11973*
*E-mail* poc@bnl.gov

ABSTRACT: Flat-field images with thick, fully-depleted CCDs exhibit response variations near the edges of the chip and at other locations, such as the regoins bordering mid-frame blooming stop implants. Two possible origins for these repsonse variations have been suggested: either photometric response (quantum efficiency) or effective pixel area is modified in these regions. In the latter case source position and shape distortions would be expected in these regions, with consequent impact on astrometric and weak lensing measurements. As an experimental check to distinguish between the two effects and to gauge the magnitude of distortion, we performed a measurment scanning an artificial star image across the affected region of one device.

# Contents

## 1. Background

Thick, fully-depleted CCDs (FDCCDs) have been selected for current and upcoming optical-NIR surveys. In these sensors the response to flat-field illumination is typically found to be nonuniform; some examples from an LSST prototype device are shown in Fig. 1. In this work we studied the ~ 30% rolloff in edge response and the more complicated structure straddling the mid-frame anti-blooming stop. Sensitivity of both effects to conditions during integration has been measured qualitatively: they are less prominent with an increase in the reverse substrate bias, a decrease in the guard drain voltage, or an increase in the illumination wavelength [1].

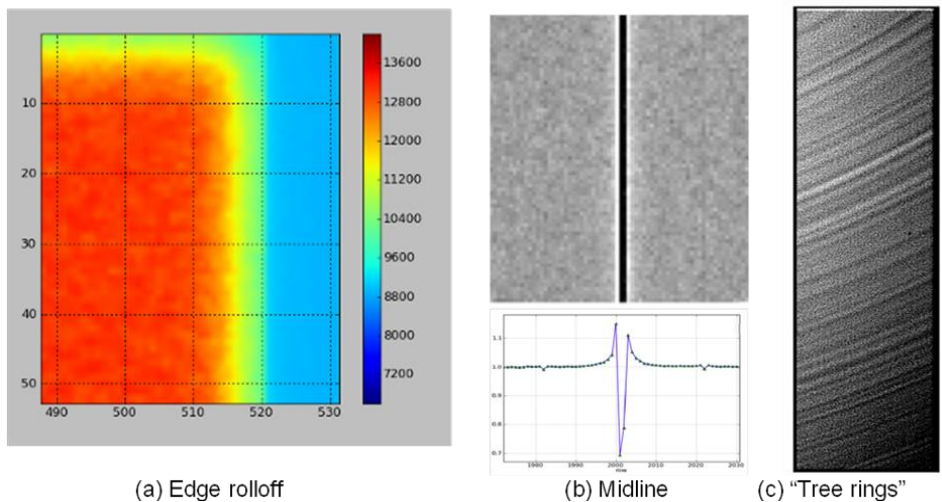

(a) Edge rolloff      (b) Midline      (c) "Tree rings"

**Figure 1 Flat-field anomalies seen in LSST prototype FDCCDs.**

## 2. Experimental detail

A prototype science-grade CCD250 device, serial number 112-01 manufactured by e2v, was studied in the CCD characterization laboratory at BNL. The geometry of the device is illustrated in Fig. 2. The illumination source was an artificial star produced by an f/1.8 spot projector [2] with a fiber-coupled 635nm laser source mounted on an automated x-y-z stage (Aerotech 3200). The chip is mounted in an LN2-cooled Dewar and is oriented such that the rows are rotated approximately 22° to the stage y-travel direction. Bias conditions were as shown in Table 1. The CCD is read out using a modified Reflex controller [3] at 545kpix/s with 118 overscan columns and 98 overscan rows.

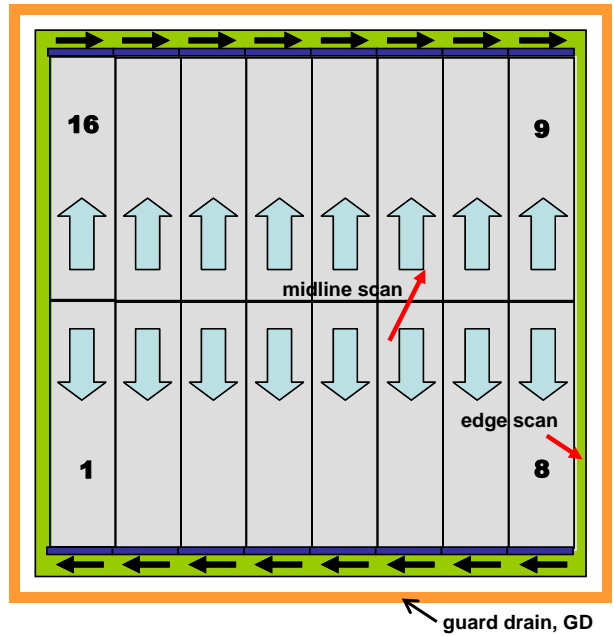

**Figure 2 CCD250 geometry. There are 16 independently-read out segments, each having 2002 imaging rows, 512 imaging columns, and 10 prescan columns. Vertical and horizontal transfer directions are indicated by arrows, and segment numbering is shown. This experiment was conducted using a spot scanned across the edge of the chip in segment 8 and across the midline between segments 6 and 11, as shown.**

All apparatus (stage, controller, laser) are under control of the RTS-2 [4] software program running on a linux PC. RTS-2 records all relevant apparatus data (including stage encoder coordinates) in the FITS header of the Primary HDU.

**Table 1 CCD operating conditions.**

| Temperature | -120C |
|---|---|
| Substrate bias | -50V |
| Guard drain bias | +26V |
| Parallel clock | 0, +8 |
| Serial clock | +0.5, +9 |
| Reset Gate | 0, +12V |
| Output gate | +2V |
| Output drain | +27V, 2mA/channel |
| Reset drain | +17V |

Prior to the SpotScan experiment we collected flat-field and bias images. For the SpotScans, the stage z-axis is first adjusted to give best focus. The resulting spot has approximately 2 pixels (20μm) FWHM with a central pixel intensity of about 19,000 electrons. Image profiles of the artificial star are shown in Fig. 3 (linear and log stretch). For the edge scan the spot was positioned at a starting coordinate in segment 8 and the stage y-axis was programmed to step in 10 μm increments. For the midline scan the initial position was in segment 6 (a non-edge segment) and the spot was scanned in the stage x-direction until it crossed into the adjoining upper segment. At each stage position two 20ms images were taken. Images were stored in multi-extension FITS format. For subsequent analysis only the image extensions 8 (edge), and 6 and 11 (midline) were used.

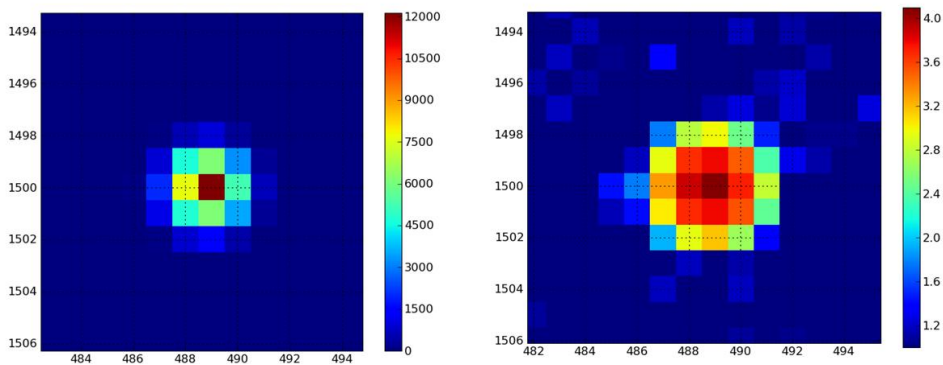

**Figure 3 Example spot profiles (ADU) shown in linear (left) and log (right) scales.**

## 3. Preliminary analysis

Stage coordinate information was extracted from the FITS headers, and then the bias-subtracted images were analyzed using SExtractor [5] with DETECT_THRESHOLD set to 10 times noise. No flat-fielding was done, as the experiment was intended to distinguish charge redistribution from photometric effects. The resulting catalogs contained centroid coordinates for the artificial star as well as flux and shape parameters. No further PSF fitting procedures were applied in this analysis.

## 4. Results

### 4.1 Edge scan

#### 4.1.1 Centroid position

Figure 4 is a composite showing the spot image at every fifth stage position. Note that the last imaging column is column 521; the rightmost image shows a partial spot PSF where the centroid has moved outside of the edge of the imaging area.

– 3 –

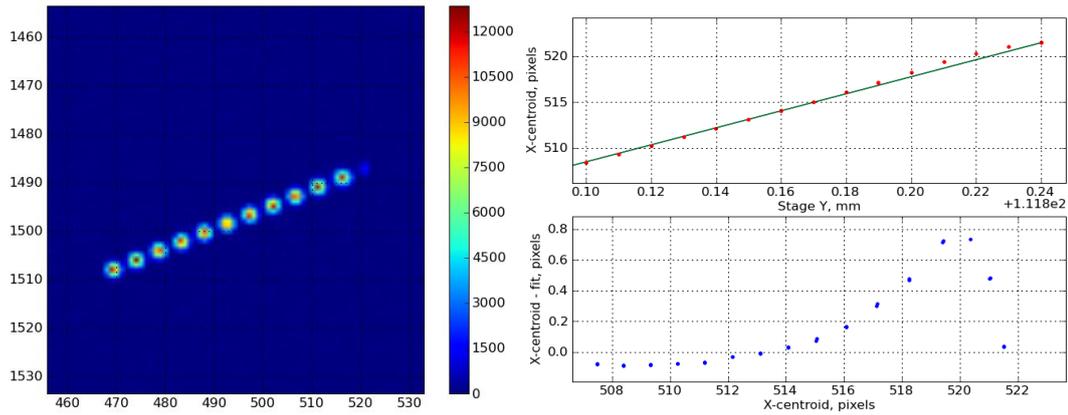

**Figure 4** 4 Left: composite showing spot images at every 5th stage position. Column 521 is the last imaging column. Right: centroid position vs. stage position and residuals.

All the following results are from the SExtractor analysis. Image centroid coordinates found by SExtractor are reported in (fractional) pixels. The x-centroid coordinate is along the row direction; columns with more positive x are closer to the chip boundary in this segment and column 521 is the last imaging column. The upper right side of Fig. 4 shows spot centroid vs. stage position in the near-edge region, and on the lower right the residuals are plotted with the x-centroid pixel coordinates on the horizontal axis. In the region starting around 8 columns (80μm) from the boundary, the position of the spot centroid is shifted in the positive-x direction relative to the expected position of the light spot as extrapolated from the linear fit in the interior.

### 4.1.2 Flux

Figure 5 shows the total flux in the spot image as a function of distance from the edge. There is no decrease in measured flux until the spot centroid is within 30μm of the edge. Because the spot profile has ~20μm FWHM, at that point the wings of the PSF have moved off the imaging area and so this should be considered a geometric, rather than a photometric effect.

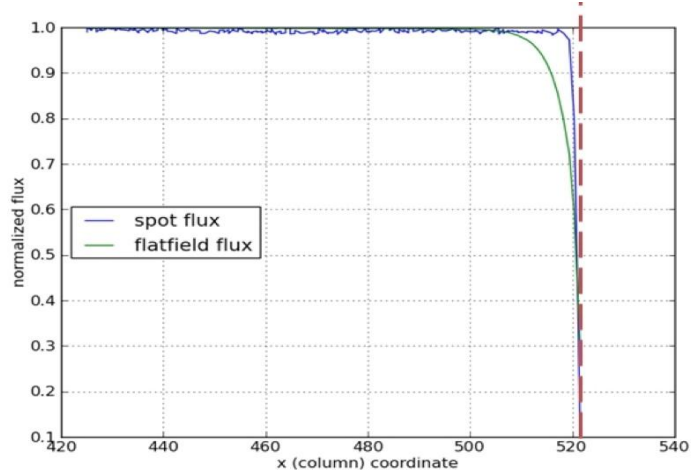

**Figure 5** Spot flux (blue) and flat-field intensity (green) vs. distance from edge.

Figure 5 also plots the flat-field intensity in the same region of the device. If the edge rolloff were due to local variations in light sensitivity, then the flux in spot images should have the same dependence on position as the measured flat-field profile. The discrepancy between flat-field and spot flux is evidence that the edge rolloff is a pixel area, or charge redistribution effect rather than a variation in photometric response, and it indicates that conventional flat-fielding would lead to photometric errors.

### 4.1.3 Shape parameters

Figure 6 shows the semi-major and semi-minor axes of the spot image as a function of edge distance. As the spot enters the rolloff region, it becomes elliptical with the major axis oriented along the x-direction. When the spot profile begins to extend beyond the edge of the imaging area, part of the spot is occulted and the shape parameters reflect this geometric cutoff.

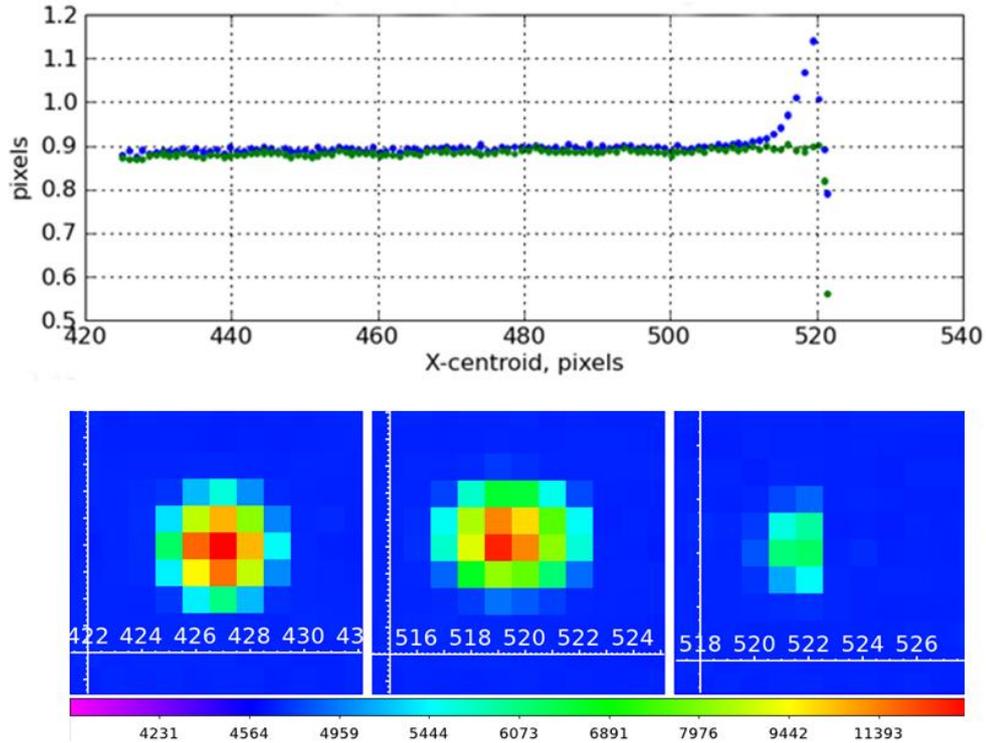

**Figure 6 Top: Semi-major (blue) and semi-minor (green) axes of the spot image as a function of edge distance. Bottom: Spot images at centroid positions 427, 519, and 522 pixels. Extracted ellipse parameters reflect the change of shape from circular to elliptical with position angle of 0; at the extreme edge of the imaging area the spot is partially occulted.**

### 4.2 Midline scan: centroid position, flux, and shape

Figure 7(a) is a composite of the images at every fourth stage position. The horizontal dashed-line shows the position of the midline – pixels below the line belong to segment 6 and pixels above the line to segment 11. In Fig. 7(b), the residuals of the centroid position are plotted as a function of stage position near the midline. Red dashed-lines are the centers of the rows adjacent to the midline. The centroid shift on either side of the midline is consistent with a repulsive lateral field displacing charge away from the blooming stop in either direction, with a region of influence of about 5 pixels (50μm).

Figure 7(c) shows the flux in the spot as a function of centroid position. Red lines are the centers of rows adjacent to midline. There is no significant effect of the midline anomaly on photometric response to the spot.

Figure 7(d) shows the FWHM of the spot versus centroid position, with red lines marking the rows adjacent to the midline. There is no trend of ellipticity with stage position. Because SExtractor does not perform PSF-fitting in its astrometry algorithms [5], the extracted centroid and FWHM of these undersampled, low S/N spot images are subject to well-known artifacts [6] which give rise to the periodic modulation and apparent large FWHM seen in the two points near row 2005.

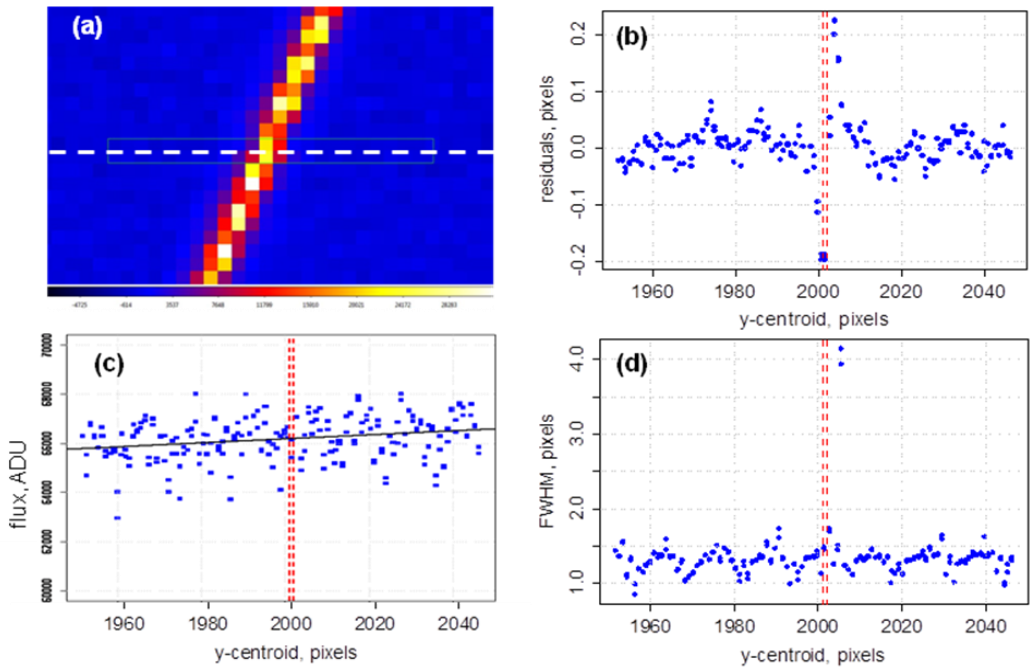

**Figure 7 (a) Composite showing spot image every fourth stage position. Horizontal line indicates chip midline location of the blooming stop. Pixels above the line shift up, below the line shift down during parallel transfer. (b) Residuals from linear fit of centroid vs. stage position. Red lines mark the centers of the rows in segment 6 and 11 adjacent to the midline. (c) Flux versus spot position. Trend line is due to illumination gradient. (d) FWHM vs. position.**

## 5. Conclusions

Scanned images of an artificial star across regions exhibiting flat-field nonuniformities show that (1) charge centroids are shifted towards regions of suppressed flat-field response and vice versa; (2) artificial star flux varies much less than flat-field flux, and (3) PSF becomes elliptical in regions of strong flat-field response variation. These effects are consistent with lateral electric fields present in the thick FDCCD that add vectorially with the vertical overdepletion field to divert photogenerated charge from drifting in a direction normal to the entrance window. Conventional image processing treatment of flat-fielding will result in photometric, astrometric, and shape measurement errors unless lateral field effects are properly taken into account.

**Acknowledgments**

Ivan Kotov, Dajun Huang Peter Doherty, and Petr Kubanek were responsible for the electronics and software used in BNL's characterization laboratory. Peter Takacs and Justine Haupt designed and constructed the optomechanics. P. Antilogus, P. Astier, and C. Stubbs are thanked for helpful discussions.

This manuscript has been co-authored by employees of Brookhaven Science Associates, LLC. Portions of this work are supported by the Department of Energy under contract DE-AC02-98CH10886 with Brookhaven National Laboratory. LSST project activities are supported in part by the National Science Foundation through Governing Cooperative Agreement 0809409 managed by the Association of Universities for Research in Astronomy (AURA), and the Department of Energy under contract DE-AC02-76-SFO0515 with the SLAC National Accelerator Laboratory. Additional LSST funding comes from private donations, grants to universities, and in-kind support from LSSTC Institutional Members.